\documentclass[aps,prl,reprint,superscriptaddress,amsmath,amssymb]{revtex4-2}
\pdfoutput=1
\usepackage{graphicx}
\usepackage{bm}
\usepackage[colorlinks=true, allcolors=blue]{hyperref}
\usepackage{physics}
\begin{document}
%
\title{Heterogeneity and Low-Frequency Vibrations in Bidisperse Sphere Packings}
%
\author{Yusuke Hara}
\email{hara-yusuke729@g.ecc.u-tokyo.ac.jp}
\affiliation{Graduate School of Arts and Sciences, The University of Tokyo, Tokyo 153-8902, Japan}
\author{Hideyuki Mizuno}
\affiliation{Graduate School of Arts and Sciences, The University of Tokyo, Tokyo 153-8902, Japan}
\author{Atsushi Ikeda}
\affiliation{Graduate School of Arts and Sciences, The University of Tokyo, Tokyo 153-8902, Japan}
\affiliation{Research Center for Complex Systems Biology, Universal Biology Institute, The University of Tokyo, Tokyo 153-8902, Japan}
%
\date{\today}
%
\begin{abstract}
In the jamming transition of monodisperse packings, spatial heterogeneity is irrelevant as the transition is described by mean-field theories. 
Here, we show that this situation drastically changes if the particle-size dispersity is large enough. 
We use computer simulations to study the structural and vibrational properties of bidisperse sphere packings with a large size ratio. 
Near the critical point, the small particles tend to form clusters, leading to the emergence of large-scale structural heterogeneity. 
Concomitantly, the low-frequency vibrations are significantly enhanced compared to those in monodisperse packings, and their density of states follows a linear law with the frequency.
We numerically and theoretically demonstrate that these behaviors of the structural heterogeneity and the low-frequency vibrations are intimately connected.  
The present work suggests that the nature of heterogeneous packings is markedly different from that of homogeneous packings. 
\end{abstract}
%
\maketitle
%
\textit{Introduction.}---
Soft matters composed of dense macroscopic particles, such as foams, emulsions, and granular materials, commonly undergo the jamming transition into disordered solid states~\cite{OHern2003,VanHecke2010}. 
This transition has attracted much attention in fundamental physics as it is an unusual critical phenomenon in non-thermal systems~\cite{OHern2003,VanHecke2010,parisi2020theory}, and it also gives a fresh insight into thermal, structural glasses~\cite{Liu1998,Wyart2005an}. 
Additionally, dense and active biological systems, such as cell tisues~\cite{Bi2015} and cytoplasm~\cite{Parry2014-xy,Nishizawa2017}, have been studied concerning the jamming transition. 

Recent dramatic advances in jamming studies have benefitted from the simplest model, composed of nearly monodisperse, frictionless spheres. 
It is now established that \textit{mean-field} theories can describe the jamming transition. 
At the transition point, a majority ($\sim 95 \%$) of particles in the system get frozen simultaneously, and the system becomes isostatic; the contact number per particle becomes $Z=2d$ where $d$ is the spatial dimension~\cite{OHern2003}.
In the jammed phase, vibrational properties are affected by the criticality of transition. 
Spatially disordered vibrations become abundant, which can not be described by the classical Debye theory based on phonon vibrations.
The vibrational density of states~(vDOS) shows a plateau in the frequency range down to the characteristic frequency $\omega_\ast$~\cite{Silbert2005}. 
At lower frequencies $\omega < \omega_\ast$, the vDOS follows the scaling law $D(\omega) \sim (\omega/\omega_\ast)^2$, which is called the non-Debye scaling~\cite{Charbonneau2016,Shimada2020a}.
These vibrational properties are explained by a replica theory for spherical particles~\cite{parisi2020theory, Franz2015} and an effective medium theory for spring networks~\cite{Wyart2005,Wyart2010,DeGiuli2014}.
These theories are mean-field, ignoring a possible \textit{spatial heterogeneity}: The former is constructed for the infinite-dimensional limit, where any spatial correlation does not exist~\cite{parisi2020theory}, while the latter disregards a heterogeneity in the distribution of local contact numbers~\cite{DeGiuli2014}. 
Therefore, the success of these theories implies that spatial heterogeneity in structure is irrelevant to the jamming transition. 
Instead, an unusual homogeneity, the hyperuniformity, in the density field~\cite{Donev2005,Ikeda2017} and the contact number distribution~\cite{Hexner2018,Hexner2019} has been debated.

\begin{figure}[t]
\centering
\includegraphics[width=0.95\linewidth]{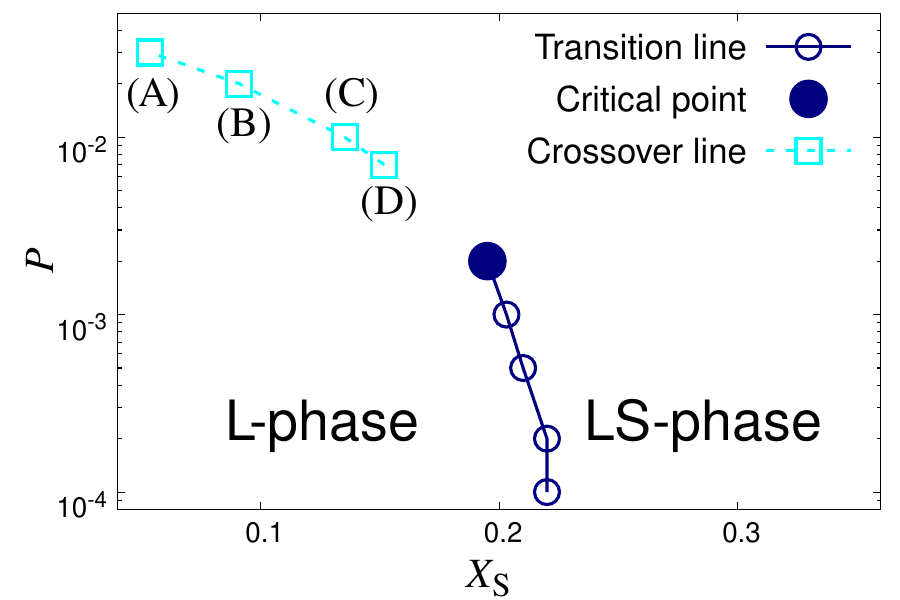}
\caption{
Phase diagram of bidisperse packings with a size ratio of 6, in the plane of the pressure $P$ and the fraction of small particles $X_{\rm S}$~\cite{Hara2021}. 
The solid line indicates the first-order transition line which terminates at the critical point. 
The dashed line indicates the crossover line where half of the small particles are jammed. 
The squares indicate the state points (A-D) on which we focus in the present work. 
} \label{diagram}
\end{figure}

In this Letter, we demonstrate that this situation drastically changes if the size dispersity of particles is large enough: Large-scale structural heterogeneity emerges, altering the low-frequency vibrations significantly.
We study a binary mixture composed of large and small particles with a large size ratio. 
Previous studies~\cite{Prasad2017, Petit2020, Hara2021, Srivastava2021, Petit2023} have established the phase diagram of such systems, as shown in Fig.~1. 
There are two distinct jammed phases: The ``L-phase" where only large particles are jammed, and the ``LS-phase" where both small and large particles are jammed. 
These phases are separated by the first-order transition line which terminates at the critical point. 
Crossing the transition line, most small particles get jammed simultaneously such that the fraction of jammed small particles changes discontinuously, whereas above the critical point, it changes continuously.

In this work, we first show that the packing structure becomes highly heterogeneous in approaching the critical point. 
The small particles tend to form clusters, and a system-spanning cluster emerges at the critical point. 
We next show that the vDOS at low-frequency regime follows a linear scaling law $D(\omega) \sim \omega/\omega_\ast$ near the critical point; the low-frequency vibrations are significantly intensified. 
Finally, we numerically and theoretically demonstrate that these two properties, large-scale heterogeneity in structure and abundance of low-frequency vibrations, are closely related. 
The present work realizes heterogeneously jammed states, which have markedly different properties from homogeneously jammed states.

\textit{Model.}---
We consider $N_{\rm S}$ small and $N_{\rm L}$ large spherical particles in a three-dimensional~($d=3$) box of linear dimension $L$.
We set the size ratio to be $\sigma_{\rm L}/\sigma_{\rm S} = 6$, where $\sigma_{\rm S}$ and $\sigma_{\rm L}$ are diameters of small and large particles, respectively. 
Particles interact via the harmonic potential $v_{ij}(r_{ij}) =\frac{\epsilon}{2}(\sigma_{ij} - r_{ij})^2 \Theta (\sigma_{ij} - r_{ij})$, where $r_{ij}$ is distance between particles $i$ and $j$, $\sigma_{ij} = (\sigma_{i}+\sigma_{j})/2$, and $\Theta(r)$ is Heaviside step function. 
Length, time, and energy are measured in units of $\sigma_{\rm S}$, $\sqrt{m_{\rm S}\sigma_{\rm S}^2/\epsilon}$, and $\epsilon$, respectively~($m_{\rm S}$ is the mass of small particles).

The macroscopic state of the system is specified by two state-variables $(P,X_{\rm S})$: Pressure $P = L^{-3} \sum_{\expval{ij}} f_{ij} r_{ij}$ with contact force $f_{ij} = v_{ij}^{\prime}(r_{ij})$~($\prime$ denotes derivative), and fraction of small particles $X_{\rm S} = N_{\rm S}/(N_{\rm S} + 6^3 N_{\rm L})$.
We fix $N_{\rm S}=8000$ and tune $N_{\rm L}$ to control $X_{\rm S}$. 
To obtain mechanically stable packings at a given $(P,X_{\rm S})$, we quenched random configurations by using FIRE algorithm~\cite{FIRE}, and tuned the system size $L$ by an iterative compression/decompression algorithm to realize the target $P$~\cite{Hara2021}. 
We prepared $200$ samples at each $(P,X_{\rm S})$ and calculated the sample average for various quantities, which is denoted by $\expval{\cdot}$. 

The jammed particles are identified as particles making at least $d+1$ contacts, as usually employed~\cite{OHern2003}. 
The fraction of the jammed small particles,
\begin{eqnarray}
R_{\rm S}=\frac{1}{N_{\rm S}} \sum_{i \in {\rm S}} S_i,
\end{eqnarray}
is an order parameter to determine the phase behavior, where ${\rm S}$ denotes a set of small particles, and $S_i = 1$ for the jammed state while $S_i=0$ otherwise. 
To systematically approach the critical point, we focus on four state points as indicated in Fig.~\ref{diagram}: (A) $(P,X_{\rm S}) = (3 \times 10^{-2},0.0536)$, (B) $(2 \times 10^{-2},0.0910)$, (C) $(10^{-2},0.135)$, and (D) $(7 \times 10^{-3},0.152)$. 
These states are located on the crossover line where half of the small particles are jammed while the other half are unjammed, \textit{i.e.}, $\expval{R_S} = 0.5$. 

\begin{figure}[t]
\centering
\includegraphics[width=0.95\linewidth]{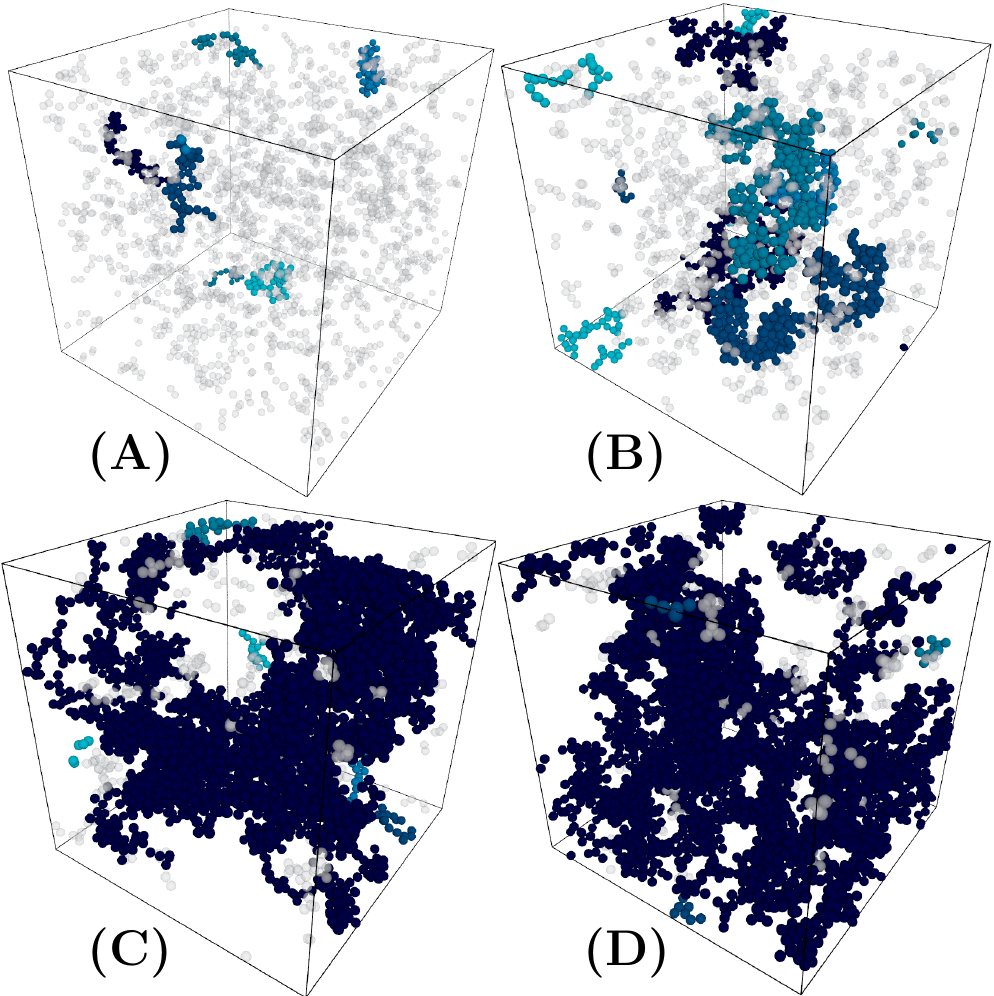}
\caption{
Typical configurations of jammed small particles at A (left top), B (right top), C (left bottom), and D (right bottom). 
The five largest clusters are in blue, and the smaller clusters are in gray. 
}
\end{figure}

{\it Structure.}---
Since the phase behavior is controlled by the jammed small particles, we first focus on their spatial structures. 
We define a cluster as a group of particles that are connected via contacts between the jammed small particles only. 
We numerically identified such clusters using Hoshen-Kopelman algorithm~\cite{Hoshen1976}.
In Fig.~2, we show snapshots of representative packings, where the five largest clusters are in blue while the other smaller clusters are in gray. 
Far from the critical point (state A), small clusters are scattered throughout the system.
In approaching the critical point (state D), the size of clusters grows so that the largest cluster spans the entire system. 
To quantify this percolation behavior, we calculated the number of particles in the largest cluster, $\expval{S_{\text{max}}}$, and showed an increasing~(most likely diverging) behavior of $\expval{S_{\text{max}}}$~[see the Supplemental Material~(SM)].

Figure~2 shows that the jammed small particles are distributed heterogeneously near the critical point. 
To discuss this heterogeneity more quantitatively, we introduce two types of radial distribution functions:
\begin{eqnarray}
&& G(r) = \frac{\expval{L^3}}{N_{\rm S}^2} \expval{\sum_{i,j \in {\rm S}} \delta (\vec{r} - \vec{r}_{ij})}, \\
&& G_{\rm S}(r) = \frac{\expval{L^3}}{\expval{\sum_{i \in {\rm S}} S_i}^2} \expval{\sum_{i,j \in {\rm S}} S_i S_j \delta (\vec{r} - \vec{r}_{ij})}, 
\end{eqnarray}
where $\vec{r}_i$ is the position vector of particle $i$, $\vec{r}_{ij}=\vec{r}_i-\vec{r}_j$, and $r = |\vec{r}|$.
$G(r)$ is a distribution function for all the small particles including unjammed ones, while $G_{\rm S}(r)$ is for the jammed small particles only. 
As shown in Fig.~3, $G(r)$ is almost identical at the states A and D, meaning that the spatial distribution of all the small particles is insensitive to the criticality. 
In contrast, $G_{\rm S}(r)$ varies near the critical point.
Far from the critical point (state A), $G_{\rm S}(r)$ becomes larger than $G(r)$ at $r \lesssim 6$, whereas $G_{\rm S}(r) \approx G(r)$ at $r \gtrsim 6$. 
Notice that the length $6$ is the diameter of the large particle, which should be in the same order as the size of pores formed by the jammed large particles. 
In approaching the critical point (state D), $G_{\rm S}(r)$ becomes substantially larger than $G(r)$, and finite spatial correlation persists even at $r \gtrsim 6$. 
These results indicate that the jamming of small particles takes place cooperatively in the entire system near the critical point, whereas far from the critical point the cooperativity persists only within each pore. 
Thus, large-scale heterogeneity emerges in the spatial distribution of the jammed small particles near the critical point. 
We emphasize that such heterogeneity is totally absent in the monodisperse packings. 

\begin{figure}[t]
\centering
\includegraphics[width=0.95\linewidth]{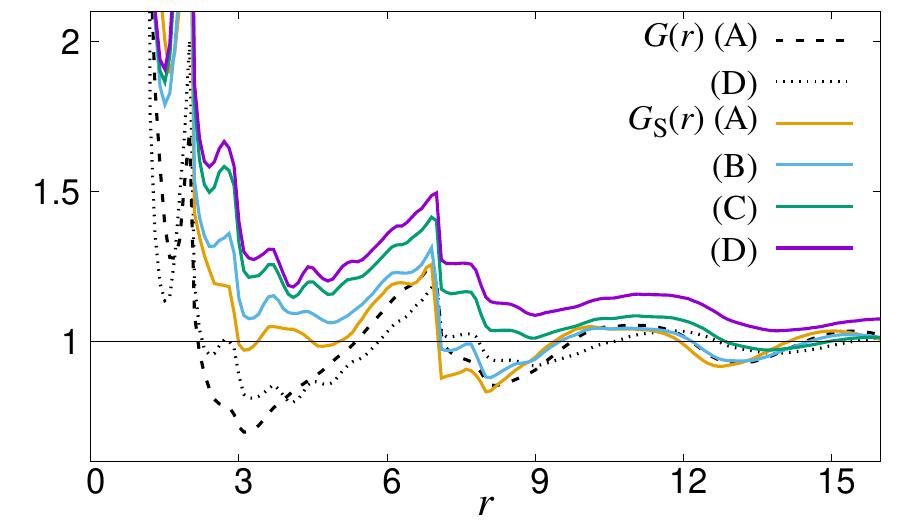}
\caption{
Radial distribution functions of the small particles.
$G(r)$ is for all the small particles in the system, while $G_{\rm S}(r)$ is for the jammed small particles only. 
}
\end{figure}

{\it Vibrations.}---
We next study the low-frequency vibrations in bidisperse packings using a standard vibrational-mode analysis~\cite{MizunoIkeda2022}. 
Vibrational modes are obtained as eigenvectors of the dynamical matrix $\mathcal{M}$.
Off-diagonal elements of $\mathcal{M}$ from different particles $i \ne j$ are
\begin{equation}
\mathcal{M}_{ij} = - k_{ij} \frac{\vec{r}_{ij}\bigotimes \vec{r}_{ij}}{r_{ij}^2} 
+ \frac{f_{ij}}{r_{ij}} \left( I - \frac{\vec{r}_{ij}\bigotimes \vec{r}_{ij}}{r_{ij}^2} \right), \label{eq:DM}
\end{equation}
where $I$ denotes unit tensor, and $k_{ij} = v_{ij}^{\prime \prime}(r_{ij})$.
Diagonal elements from same particle $i=j$ are $\mathcal{M}_{ii} = - \sum_{j \in \partial_i} \mathcal{M}_{ij}$, where $\partial_i$ indicates a set of particles contacting with the particle $i$. 
The vDOS is then calculated as $D(\omega)=\frac{1}{3 N } \langle \sum_{k} \delta(\omega - \omega_k) \rangle$,  where $\omega_k$ the $k$-th eigenfrequency.

\begin{figure}[t]
\centering
\includegraphics[width=0.95\linewidth]{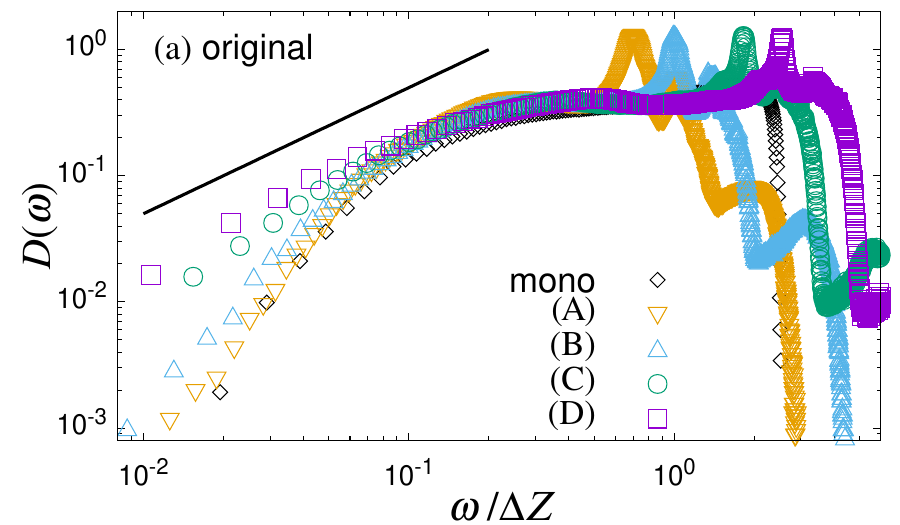}
\includegraphics[width=0.95\linewidth]{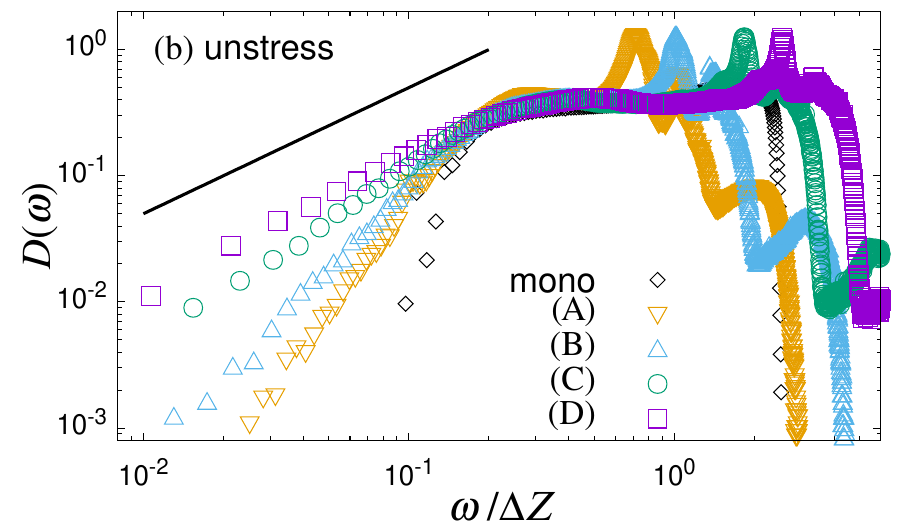}
\caption{
The vDOSs of the bidisperse packings at the states (A-D) and that of the monodisperse packing at $P=7 \times 10^{-3}$ for (a) the original system and (b) the unstressed system.
The frequency $\omega$ is scaled by the contact number $\Delta Z$.
The black line represents the linear dependence on $\omega$, \textit{i.e.}, $D(\omega) \propto \omega$.
}
\end{figure}

Figure~4~(a) shows the vDOSs of the bidisperse packings, compared to that of the monodisperse packing. 
The frequency $\omega$ is scaled by the excess contact number $\Delta Z = Z -2d$ since the characteristic frequency follows $\omega_\ast \propto \Delta Z$ in the monodisperse packings~\cite{Silbert2005}.
In the high-frequency regime $\omega /\Delta Z \gtrsim 10^{-1}$, the vDOSs of all cases collapse well and show a plateau. 
This result indicates that the critical point in the bidisperse packings does not affect anomalous vibrational modes in the plateau regime at $\omega > \omega_\ast$.
In contrast, in the low-frequency regime $\omega /\Delta Z \lesssim 10^{-1}$, $D(\omega)$ strongly depends on the state; $D(\omega)$ gets larger in approaching the critical point, and near the critical point, it shows the linear dependence on $\omega$ as $D(\omega) \sim \omega/\omega_\ast$.
This linear dependence is in sharp contrast to the monodisperse packings which show the non-Debye scaling law for the boson peak, $D(\omega) \sim (\omega/\omega_\ast)^2$, and the quartic law for the quasi-localized vibrational modes, $D(\omega) \sim (\omega/\omega_\ast)^4$~\cite{Mizuno2017}.
We also calculated the ``concentration" on small particles $M^S_k$ for these modes (see the SM).
A remarkable feature is that in the regime of $D(\omega) \sim \omega/\omega_\ast$, vibrational modes concentrate on small particles only, while the contribution from large particles is negligible.
Our results establish that the low-frequency vibrations in near-critical bidisperse packings are significantly intensified compared to monodisperse packings, which are dominated by small particles.

In the monodisperse packings, the non-Debye scaling $D(\omega) \sim (\omega/\omega_\ast)^2$ is a consequence of the marginal stability~\cite{Franz2015,DeGiuli2014}. 
This can be numerically confirmed by analyzing the unstressed version of the system~\cite{Wyart2005}.
In this analysis, the contact force $f_{ij}$ in Eq.~(\ref{eq:DM}) is artificially set to be zero, and we diagonalize the resulting dynamical matrix. 
Since the contact forces always destabilize the system, this artificial operation improves the stability of the packings.
As a result, the low-frequency vibrations (except for phonon vibrations) at $\omega \lesssim \omega_\ast$ disappear in the unstressed system~\cite{Wyart2005,Xu2010,Mizuno2017,Lerner2018}. 

Here, we analyze the unstressed version of the bidisperse packings and the monodisperse packing, and show their vDOSs in Fig.~4~(b). 
In the monodisperse packing, $D(\omega)$ at $\omega \lesssim \omega_\ast$ sharply drops as expected.
On the contrary, such a drop does not appear for the bidisperse packings, and the vDOS near the critical point still follows the linear law $D(\omega) \sim \omega/\omega_\ast$.
This observation indicates that the origin of the low-frequency vibrations in bidisperse packings differs fundamentally from monodisperse packings: The low-frequency vibrations in bidisperse packings do \textit{not} originate from the instability due to the contact forces.
Instead, as shown below, they are explained by the heterogeneities in the structures.
In the SM, we also show that the participation ratio $P_k$ of each mode $k$ is unchanged between original and unstressed systems, which further validates the above discussion.

\textit{Link between structure and vibrations.}---
Now, we will demonstrate that the abundance of low-frequency vibrations is linked to structural heterogeneity. 

We start with numerical evidence that spatial distributions of local contact numbers and low-frequency vibrations are correlated.  
We define the ``concentration" of the low-frequency vibrations on the jammed small particle $i$,
\begin{equation}
M_i = \sum_{\omega_k < \omega^*} \abs{\vec{e}_{i,k}}^2, \label{eq:Mi}
\end{equation}
where $\vec{e}_{i,k}$ is the $i$-th particle displacement in the $k$-th vibrational mode.
We next define the local contact number as follows. 
In general, a contact between two particles constrains their relative motion. 
However, if one of the particles is immobile, the contact virtually constrains the motion of the other one only.  
In the present bidisperse packings, the low-frequency vibrations are concentrated on the jammed small particles, meaning that the large particles virtually act as such immobile obstacles.
To take this into account, we introduce an effective contact number for the jammed small particle $i$,
\begin{equation}
Z_i = Z^{SS}_i + 2 Z^{SL}_i, \label{eq:Zi}
\end{equation}
where $Z^{SS}_i$ and $Z^{SL}_i$ are the number of contacts with small and large particles, respectively. 
Since the large particles are immobile, the isostatic condition is that the average of this effective contact number is equal to $2d$.

We computed above $M_i$ and $Z_i$, which are then coarse-grained as $\bar{Z}_i = \frac{Z_i + \sum_{j \in \partial i} Z_j}{1 + N_{\partial i}}$ and $\bar{M}_i = \frac{M_i + \sum_{j \in \partial i} M_j}{1 + N_{\partial i}}$ where $N_{\partial i}$ is the number of neibouring particles for the particle $i$. 
To look at correlations between $\bar{Z}_i$ and $\bar{M}_i$, we calculated a conditional distribution function, $\text{P}(\bar{Z} | \bar{M}) = \langle \sum_{i \in {\rm S}} \delta(\bar{M}_{i} - \bar{M}) \delta (\bar{Z}_i - \bar{Z}) \rangle / \langle \sum_{i \in {\rm S}} \delta(\bar{M}_i - \bar{M}) \rangle$. 
Figure~5 shows $\text{P}(\bar{Z} | \bar{M})$ near the critical point (state D). 
The distribution shifts to the left with increasing $\bar{M}$: The more a particle participates in the low-frequency vibrations, the smaller its local contact number is. 
For the largest $\bar{M}$, the distribution peaks at $\bar{Z} \approx 5$ which is even smaller than $2d~(=6)$. 
This result clearly demonstrates that particles with fewer contact numbers contribute more to low-frequency vibrations.

\begin{figure}[t]
\centering
\includegraphics[width=0.95\linewidth]{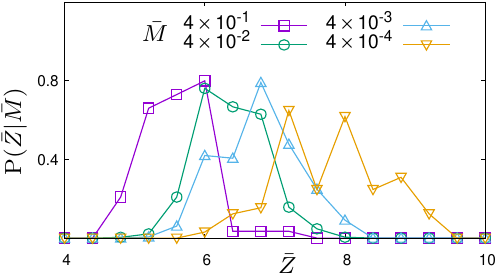}
\caption{Conditional distribution function $\text{P}(\bar{Z} | \bar{M})$ for the bidisperse packing near the critical point~(state D). 
$\text{P}(\bar{Z} | \bar{M})$ for $\bar{M}=4 \times 10^{-n}$ with $n=1,2,3,4$ is plotted as a function of $\bar{Z}$. 
}
\end{figure}

The above numerical evidence motivates us to explain the linear law $D(\omega) \sim \omega/\omega_\ast$ by considering the local contact-number fluctuations. 
Here we generalize the ``cutting argument"~\cite{Wyart2005}, which was originally proposed to explain the plateau in the vDOS for monodisperse packings.
We consider a system with the excess contact number $\Delta Z$ and linear dimension $L$, and divide it into $(L/\ell)^3$ subsystems of linear dimension $\ell$.
In the subsystem $n$~($n=1,2,...,(L/\ell)^3$) with the excess contact number $\delta z_n$, we estimate the number of the soft modes as $\mathcal{N}_{{\rm soft},n}(\ell) = - \delta z_n \ell^3 + \ell^2$ for $\delta z_n < 1/\ell$ while $\mathcal{N}_{{\rm soft},n}(\ell) = 0$ otherwise, where unimportant $\mathcal{O}(1)$ coefficients are omitted. 
Since a soft mode in each subsystem can be used to generate a trial mode in the original system with the frequency $\omega \sim 1/\ell$, the cumulative vDOS is bounded as $C(\omega) \gtrsim L^{-3} \sum_n \mathcal{N}_{{\rm soft},n}(\ell \sim 1/\omega)$, thus we obtain
\begin{eqnarray}
C(\omega) \gtrsim \omega \int^{\omega}_0 \text{P}_{\ell} (\delta z) \left( 1 - \frac{\delta z}{\omega} \right) d \delta z, \label{eq:Comega}
\end{eqnarray}
where $\text{P}_{\ell} (\delta z)$ is probability distribution to find subsystems with $\delta z$.
Eq.~(\ref{eq:Comega}) is a generalization of the original cutting argument, which is embedded by variation in $\delta z$ among the subsystems.
Note that in the original argument for monodisperse packings~\cite{Wyart2005}, we suppose the delta function for $\text{P}_{\ell} (\delta z) = \delta (\delta z - \Delta Z)$, which provides $D(\omega)= dC/d\omega \gtrsim \omega^0$ for $\omega > \Delta Z$.

Now we assume that spatial fluctuations in $\delta z$ are sufficiently large; their standard deviation is in the same order as the average $\Delta Z$, so a certain fraction of subsystems with $\delta z \sim 0$ exists. 
In such the case, $\text{P}_{\ell} (\delta z)$ can be virtually treated as a constant for $\omega~(\sim 1/\ell) < \omega_\ast = \Delta Z$, and the delta function $\delta (\delta z - \Delta Z)$ for $\omega \gg \omega_\ast = \Delta Z$~(see the SM for a specific example with the Gaussian distribution). 
We thus obtain $D(\omega) \gtrsim \omega/\omega_\ast$ for $\omega < \omega_\ast$ and $D(\omega) \gtrsim \omega^0$ for $\omega \gg \omega_\ast$. 
Assuming that the inequality is saturated as in the original argument, we obtain $D(\omega) \sim \omega/\omega_\ast$ for $\omega < \omega_\ast$ and $D(\omega) \sim \omega^0$ for $\omega \gg \omega_\ast$, which totally agree with our numerical results in Fig.~4.  

\textit{Concluding remark.}---
In this Letter, we investigated the structural and vibrational properties of bidisperse packings with a large size ratio.
In approaching the critical point, the jamming of small particles takes place cooperatively, leading to the emergence of large-scale structural heterogeneity.
Concomitantly, the low-frequency vibrations are significantly intensified near the critical point, and the vDOS follows the linear law, $D(\omega) \sim \omega/\omega_\ast$, at $\omega < \omega_\ast \propto \Delta Z$.
We numerically and theoretically demonstrated that these two phenomena of spatial heterogeneity and intensified low-frequency vibrations are ultimately connected.
In particular, we generalized the cutting argument to the heterogeneous packings, and successfully explained the linear law of the vDOS in terms of spatial fluctuations in the local contact numbers.

In this work, we also found that contact forces on small particles are significantly small, and interestingly, their distribution follows a power law with a negative exponent~(see the SM), which is marked contrast to a positive exponent in monodisperse packings~\cite{charbonneau2015jamming}.
It is an interesting future work to theoretically explain this behavior by extending the stability analysis~\cite{wyart2012marginal} or the microscopic replica theory~\cite{parisi2020theory}.

Finally, our results suggest that the particle-size dispersity can cause a large-scale heterogeneity, exerting a strong impact on the vibrational properties of the packings. 
Since low-frequency vibrations play important roles in plasticy~\cite{Manning2011}, it is interesting to study the non-linear mechanical properties to elucidate impacts from the low-frequency vibrations found in this work.
More broadly, as many soft jammed solids, ranging from granular materials to biophysical systems, quite usually have continuous polydispersities with wide-range distributions~\cite{sammis1987kinematics, huang2001hydrocolloids, kwok2020apollonian, shimamoto2023common}, it should be very interesting to study such systems along the line of this work.

\begin{acknowledgments}
This work was supported by Hosokawa Powder Technology Foundation Grant No. HPTF21509, 
JST SPRING Grant No.JPMJSP2108,
and JSPS KAKENHI Grants No.
20H01868, 
20H00128,
22K03543,
23H04495,
23KJ0368.
\end{acknowledgments}

\bibliography{main}

\end{document}


%
\title{Supplementary Information for \\
Heterogeneity and Low-Frequency Vibrations in Bidisperse Sphere Packings}
%
\author{Yusuke Hara}
\email{hara-yusuke729@g.ecc.u-tokyo.ac.jp}
\affiliation{Graduate school of arts and science, University of Tokyo, Komaba, Tokyo 153-8902, Japan}
\author{Hideyuki Mizuno}
\affiliation{Graduate school of arts and science, University of Tokyo, Komaba, Tokyo 153-8902, Japan}
\author{Atsushi Ikeda}
\affiliation{Graduate school of arts and science, University of Tokyo, Komaba, Tokyo 153-8902, Japan}
\affiliation{Research Center for Complex Systems Biology, Universal Biology Institute, University of Tokyo, Komaba, Tokyo 153-8902, Japan}
%
\date{\today}
%
\maketitle
%
\subsection{Largest size of clusters of jammed small particles}
\begin{figure}[h]
\centering
\includegraphics[width=0.95\linewidth]{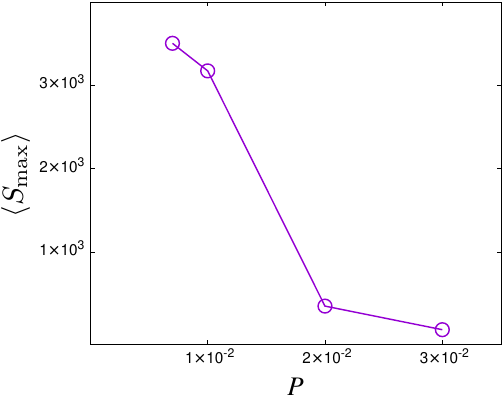}
\caption{ \label{fig-smax}
The number of particles in the largest cluster of small jammed particles, $\expval{S_{\rm max}}$, is plotted as a function of $P$ across the A, B, C, and D states.
}
\end{figure}

As mentioned in the main text, we used the Hoshen-Kopelman algorithm to identify clusters of small jammed particles.
Here, we are presenting data on the largest size of these clusters.
Figure~\ref{fig-smax} shows the plot of the number of particles in the largest cluster, $\expval{S_{\rm max}}$, as a function of the pressure $P$ across the A, B, C, and D states.
The data indicate that as the pressure decreases and the critical point is approached, $\expval{S_{\rm max}}$ increases monotonically and is most likely to diverge at the critical point.

\subsection{Low-frequency vibrational properties}
Here we are presenting supplementary information about vibrational modes.
We show the concentration on the small particles $M^S_k$ in each vibrational mode $k$.
To measure the $M^S_k$, we sum up the squared displacements $\abs{\vec{e}_{i, k}}^2$ of all the small particles $i \in \text{S}$, as
\begin{equation}
M^S_k = \sum_{i \in \text{S}} \abs{\vec{e}_{i, k}}^2.
\end{equation}
Note that the range of $M^S_k$ is between 0 and 1. 
We can also define the concentration on the large particles $M^L_k$, which is calculated by summing up $\abs{\vec{e}_{i, k}}^2$ of all the large particles $i$.
Since the vector $\vec{e}_{i, k}$ is normalized by $\sum_{i=1}^N \abs{\vec{e}_{i, k}}^2 = 1$, $M^L_k = 1-M^S_k$.

In addition, we are presenting data on the participation ratio, denoted by $P_k$, for each vibrational mode $k$~\cite{mazzacurati1996low}.
The participation ratio is a measure of the fraction of particles that contribute to the mode $k$, which is defined as
\begin{equation}
P_k = \frac{1}{N} \left[\sum_{i=1}^N \left(\vec{e}_{i, k} \cdot \vec{e}_{i, k}\right)^2\right]^{-1}.
\end{equation}
$P_k$ ranges from $1/N$ to $1$: If the mode $k$ is localized to a single particle, $P_k$ equals $1/N$, while if all the particles equally contribute to the mode $k$, then $P_k$ equals $1$.

\subsubsection{Concentration on small particles}
\begin{figure*}[t]
\includegraphics[width=.95\linewidth]{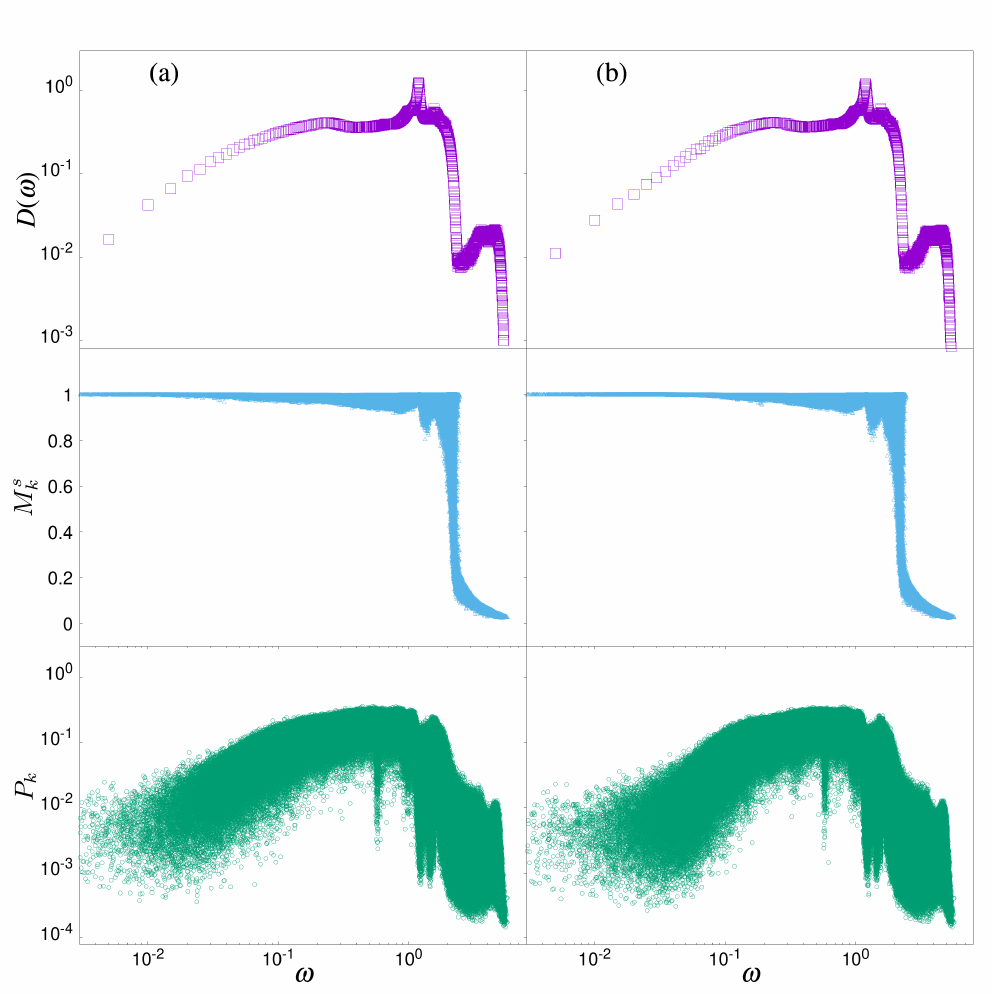}
\caption{
The vDOS $D(\omega)$, the concentration on small particles $M^S_k$, and the participation ratio $P_k$ as functions of the eigenfrequency $\omega_k$ for the state D near the critical point.
The data pertain to (a) the original system and (b) the unstressed system.
}
\label{fgr:PR}
\end{figure*}

In Fig.~\ref{fgr:PR}(a), we are presented with data on $M^S_k$ as a function of the eigenfrequency $\omega_k$, along with the vDOS $D(\omega)$, for the state D near the critical point.
The data clearly illustrates that small particles are the main contributors to the vibrational modes in the low-frequency range, as mentioned in the main text.
Specifically, in the range where the vDOS follows a linear law $D(\omega) \sim \omega/\omega_\ast$, values of $M^S_k$ are approximately equal to 1 while values of $M^L_k$ are approximately equal to 0.
This result means that small particles contribute dominantly to the low-frequency modes, while the contribution from large particles is negligible.

\subsubsection{Comparison between original system and unstressed system}
Figure~\ref{fgr:PR} presents the data for $D(\omega)$, $M^S_k$, and $P_k$ for both the original system in (a) and the unstressed system in (b). 
Upon comparing the data of (a) and (b), we observe that they show similar results in all of $D(\omega)$, $M^S_k$, and $P_k$.
This observation supports the discussion presented in the main text, which suggests that the origin of the low-frequency vibrations in bidisperse packings is fundamentally different from monodisperse packings.
Specifically, the low-frequency vibrations in bidisperse packings do not originate from the instability caused by contact forces.

\subsection{Cutting argument with Gaussian distribution of contact numbers}
In this study, we have extended the cutting argument by introducing spatial heterogeneity in the contact-number distribution.
As a result, we have derived an inequality equation for the cumulative vDOS, given by Eq.~(7) in the main text:
\begin{eqnarray}
C(\omega) \gtrsim \omega \int^{\omega}_0 \text{P}_{\ell} (\delta z) \left( 1 - \frac{\delta z}{\omega} \right) d \delta z.  \label{eq:cvdos}
\end{eqnarray}
Here, we consider a specific case where the contact-number distribution $\text{P}_{\ell} (\delta z)$ is given by the Gaussian distribution with the standard deviation equal to the average value $\Delta Z$, as described by
\begin{equation}
\text{P}_{\ell}(\delta z) = \frac{1}{\sqrt{2\pi \Delta Z^2}} \exp\left\{- \frac{\left(\delta z - \Delta Z\right)^2}{2 \Delta Z^2}\right\}. \label{eq:pdz}
\end{equation}
By substituting Eq.~(\ref{eq:pdz}) into Eq.~(\ref{eq:cvdos}), we obtain~(omitting unimportant $\mathcal{O}(1)$ coefficients)
\begin{widetext}
\begin{equation}
\frac{C(\omega)}{\Delta Z} \gtrsim \exp \left\{ -\frac{\left(\frac{\omega}{\Delta Z}-1\right)^2}{2} \right\} - \exp\left(-\frac{1}{2}\right) + \left(\frac{\omega}{\Delta Z}-1\right) \int^{\frac{\omega}{\Delta Z}}_0 \exp \left\{-\frac{(x-1)^2}{2} \right\} dx,
\end{equation}
which then provides an expression for the vDOS $D(\omega)=dC/d\omega$:
\begin{equation}
D(\omega) \gtrsim \int^{\frac{\omega}{\Delta Z}}_0 \exp\left\{ -\frac{(x-1)^2}{2} \right\} dx \sim
\begin{cases}
\frac{\omega}{\Delta Z} & \left(\omega \ll \Delta Z \right), \\
\omega^0 & \left(\omega \gg \Delta Z \right).
\end{cases}
\end{equation}
\end{widetext}
Assuming that the inequality is saturated, this theoretical prediction is consistent with our numerical observations on the vDOS.
In particular, this extended cutting argument successfully explains the linear law $D(\omega) \sim \omega/\omega_\ast$ at $\omega < \omega_\ast$.
We note that this prediction is valid as long as $\text{P}_{\ell}(\delta z) \sim 1/ \Delta Z$ in the range of $\delta z \lesssim \Delta Z$, with a rapid decay to zero in $\delta z \gtrsim \Delta Z$.

\subsection{Contact force distribution}
%


\begin{figure}[h]
\centering
\includegraphics[width=0.95\linewidth]{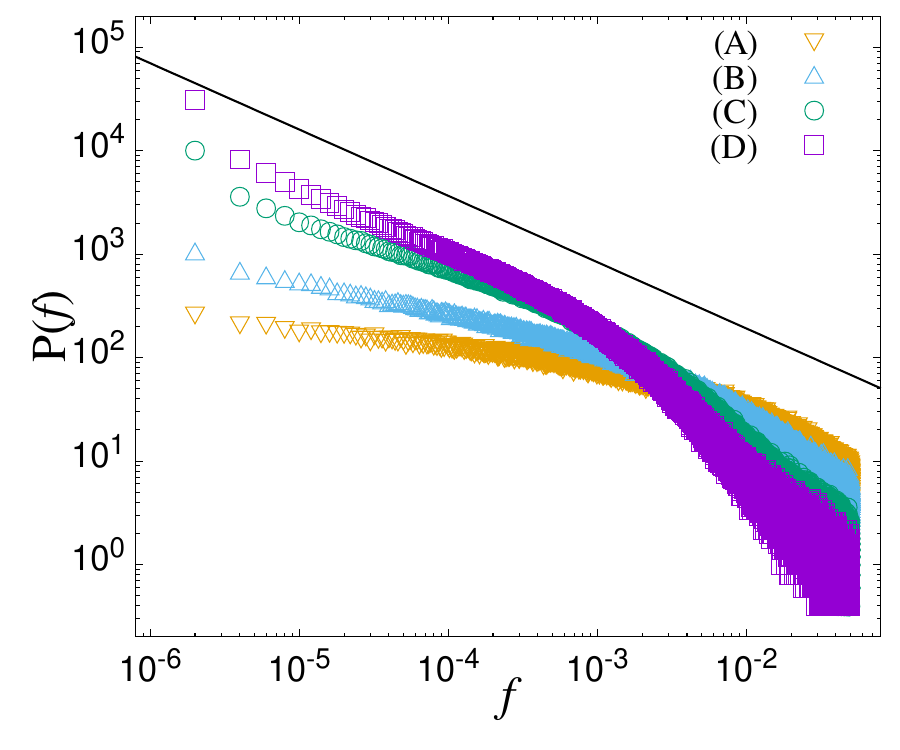}
\caption{
Contact force distribution ${\rm P}(f)$ for the A, B, C, and D states.
${\rm P}(f)$ exhibits the power-law behaviour at small $f$ with the negative exponent $\theta$.
Estimated value of $\theta$ is given in Table~\ref{table_theta}.
\label{fig:force-distribution}
}
\end{figure}


It is well understood that in monodisperse packings, the nature of marginal stability is reflected in the abundance of weak contact forces~\cite{wyart2012marginal}.
At the jamming point, the contact force distribution ${\rm P}(f)$ displays a power-law tail at small $f$, ${\rm P}(f) \propto f^{\theta}$, with a non-trivial, positive exponent $\theta~(>0)$.
The replica theory is able to accurately predict this power-law behavior with the exact value of $\theta$~\cite{charbonneau2015jamming}.

In the present work, we examine the distribution of contact forces in bidisperse packings and present results for the A, B, C, and D states in Fig.~\ref{fig:force-distribution}.
The results show the power-law behaviors at small contact forces, ${\rm P}(f) \propto f^{\theta}$.
However, the exponent $\theta$ takes negative values, in contrast to monodisperse packings where positive values are observed.
As we approach the critical point, this behavior becomes more noticeable with the larger value of $|\theta|$.
We provide the values of $\theta$ in Table~\ref{table_theta}.
It is an interesting future work to theoretically explain this remarkable behavior by extending the stability analysis~\cite{wyart2012marginal} or the microscopic replica theory~\cite{charbonneau2015jamming}.

\begin{table}[h]
 \caption{ \label{table_theta}
Exponent $\theta$ in the power-law behavior ${\rm P}(f) \propto f^{\theta}$.}
  \begin{tabular}{c cccc}
   \hline \hline
    State & A & B & C & D \\
   \hline \hline
   $\theta$ & $- 0.157$ & $- 0.269$ & $- 0.531$ & $- 0.641$ \\ 
   \hline \hline
  \end{tabular}
\end{table}



\bibliography{suppl}
\bibliographystyle{unsrtnat} 

